\documentclass[3p,preprint]{elsarticle}
\usepackage{verbatim} 
\usepackage{dsfont}
\usepackage{amsmath,amssymb}
\usepackage{epsfig}
\usepackage[active]{srcltx}
\usepackage{wrapfig}

\def\x'{\mathaccent 19 x}
\def\y'{\mathaccent 19 y}
\def\n'{\mathaccent 19 n}
\def\u'{\mathaccent 19 u}
\def\et'{\mathaccent 19 \eta}
\def\th'{\mathaccent 19 \theta}
\def\lam'{\mathaccent 19 \lambda}
\def\varet'{\mathaccent 19 \vartheta}
\def\rh'{\mathaccent 19 \rho}
\def\ph'{\mathaccent 19 \Phi}
\def\xb'{\mathaccent 19 {\bar{x}}}

\def\bsp{\be\begin{split}}

\def\be{\begin{equation}}
\def\ee{\end{equation}}
\newcommand{\slsh}[1]{{#1}\!\!\!\! / \, }
\newcommand{\bea}{\begin{eqnarray}}
\newcommand{\eea}{\end{eqnarray}}

\journal{Physics Letters B}
\begin{document}
\begin{frontmatter}
\title{New supersymmetric Wilson loops in ABJ(M) theories}
\author[Florence]{V.~Cardinali}
\ead{cardinali@fi.infn.it}
\author[Parma]{L.~Griguolo}
\ead{griguolo@fis.unipr.it}
\author[Florence]{G.~Martelloni}
\ead{martelloni@fi.infn.it}
\author[Florence]{D.~Seminara}
\ead{seminara@fi.infn.it}
\address[Florence]{Dipartimento di Fisica, Universit\`a di Firenze and INFN Sezione di Firenze,
Via G. Sansone 1, 50019 Sesto Fiorentino, Italy}
\address[Parma]{Dipartimento di Fisica, Universit\`a di Parma and INFN Gruppo Collegato di
Parma, Viale G.P. Usberti 7/A, 43100 Parma, Italy}

\begin{abstract}
We present two new families of Wilson loop operators in ${\cal N}= 6$ supersymmetric Chern-Simons theory. The first one is defined for an arbitrary contour on the three dimensional space and it resembles the Zarembo's construction in ${\cal N}=4$ SYM. The second one involves arbitrary curves on the two dimensional sphere. In both cases one can add certain scalar and fermionic couplings to the Wilson loop so it preserves at least two supercharges. Some previously known loops, notably the 1/2 BPS circle, belong to this class, but we point out more special cases which were not known before. They could provide further tests of the gauge/gravity correspondence in the ABJ(M) case and interesting observables, exactly computable by localization techniques.
\end{abstract}
\begin{keyword}
Wilson loops - Supersymmetric gauge theories - Chern-Simons-matter theories
\end{keyword}
\end{frontmatter}

\renewcommand{\thefootnote}{\arabic{footnote}}
\setcounter{footnote}{0}

\section{Introduction and results}
Three-dimensional ${\cal N}=6$ supersymmetric Chern-Simons-matter theories with gauge group $U(N)\times U(M)$ \cite{Aharony:2008ug,Aharony:2008gk} provide an exciting arena where studying the duality between string theories on asymptotically $AdS$ spaces and conformal field theories. The gravity dual of this theory is M-theory on $AdS_4\times S^7/\mathbb{Z}_k$, where $k$ is the level of the Chern-Simons term, or, for large enough $k$, type IIA string theory on $AdS_4\times \mathbb{CP}^3$.

Like in more familiar gauge theories, it is possible to define Wilson loop operators, which in the dual string theory are given by semi-classical string surfaces \cite{Rey:1998ik,Maldacena:1998im}. The most symmetric string of this type preserves half of the supercharges of the vacuum (as well as an $U(1)\times SL(2, \mathbb{R}) \times SU(3)$ bosonic symmetry) and its dual operator in the field theory has been ingeniously derived in \cite{Drukker:2009hy} (see \cite{Lee:2010hk} for an alternative derivation in terms of potential between heavy $W$-bosons). Other Wilson loop operators, previously constructed in \cite{Drukker:2008zx,Chen:2008bp,Rey:2008bh}, preserve only 1/6 of the supercharges and are therefore not viable candidates to be the dual of this classical string. The construction of the 1/2 BPS operator uses in an essential way the quiver structure of the theory. In addition to the gauge fields, the Wilson loop couples to bilinears of the scalar fields and, crucially, also to the fermionic fields transforming in the bi-fundamental representation of the two gauge groups. The operator is classified by representations of the supergroup $U(N|M)$ and is defined in terms of the holonomy of a superconnection of this supergroup: the analysis presented in \cite{Drukker:2009hy} considers loops supported along an infinite straight line and along a circle. 

For the 1/6 BPS Wilson loop a matrix model, describing its vacuum expectation value, has been derived in \cite{Kapustin:2009kz} 
and this result 
carries over to the 1/2 BPS case. The calculation of \cite{{Kapustin:2009kz}} uses localization with respect to a specific supercharge which is also shared by the 1/2 BPS operator. This Wilson loop is cohomologically equivalent to a very specific choice of the 1/6 BPS loop, constructed with bosonic couplings only,  and is therefore also given by a matrix model. Happily it can be calculated for all values of the coupling also beyond the planar approximation \cite{Marino:2009jd,Drukker:2010nc,Drukker:2011zy} and, in the strong coupling regime, it matches string computations.

In four-dimensional ${\cal N}=4$ super Yang-Mills theory the original examples of 1/2 BPS Wilson loops (the straight line and the circle \cite{Erickson:2000af,Drukker:2000rr}) can be embedded into whole families preserving between 2 and 16 supercharges. The straight line has been generalized by Zarembo  \cite{Zarembo:2002an}, the amount of conserved supersymmetry being related to the dimension of the subspace containing the contour. An interesting property of those loops is that their expectation values seem to be trivial. The circular Wilson loop, that can be computed exactly through localization \cite{Pestun:2007rz}, has been instead generalized in \cite{Drukker:2007qr} to a class of contours living in an $S^3$ (also called DGRT loops). A subset of those operators, preserving 1/8 of the original supersymmetry, are contained in a $S^2$ and their quantum behavior is described by perturbative \cite{Bassetto:1998sr} two-dimensional Yang-Mills theory \cite{Drukker:2007qr,Bassetto:2008yf,Young:2008ed,Pestun:2009nn} (a property that is also shared by loop correlators \cite{Giombi:2009ms,Bassetto:2009rt,Bassetto:2009ms}). We remark that in ${\cal N}=4$ SYM a general classification of supersymmetric Wilson loops does exist \cite{Dymarsky:2009si,Cardinali:2012sy}.

In this letter we present two new families of BPS Wilson loops operators in ABJ(M) theories, generalizing respectively the straight line and the circle constructed in \cite{Drukker:2009hy}: they can be considered the analogous of the Zarembo and DGRT loops in three dimensional ${\cal N}=6$ super Chern-Simons-matter theories. Remarkably we recover within our analysis some BPS configurations that we have introduced in \cite{Griguolo:2012iq}, where a generalized cusped Wilson loop has been carefully studied at classical and quantum level (see also \cite{Forini:2012bb} for a discussion at strong coupling). Our results might be useful in studying the connection, originally proposed in $D=4$ by \cite{Drukker:2011za}, between quark-antiquark potential and cusp anomalous dimension \cite{Drukker:2012de,Correa:2012hh}. Potentially they could also play a role in the exact computation of the elusive function $h(\lambda)$ \cite{Nishioka:2008gz,Gaiotto:2008cg,Grignani:2008is}, as suggested in \cite{Correa:2012at}.
 
We start from the Wilson loop defined as the holonomy of the super-connection introduced by Drukker and Trancanelli \cite{Drukker:2009hy}, parameterized by a certain number of path-dependent functions $M_{J}^{\ \ I}(\tau)$, $\hat M_{J}^{\ \ I}(\tau)$, $\eta_{I}^{\alpha}(\tau)$ and $\bar{\eta}^{I}_{\alpha}(\tau)$ that specify the local couplings of bosons and fermions living in ABJ(M) theory. Our strategy is to derive first a general set of algebraic and differential conditions that correspond to preserve locally a fraction of supersymmetry, up to total derivative terms along the contour. Then we have to impose that solutions of these constraints can be combined into a conformal Killing spinor, 
\be
\label{KillSpin}
\bar\Theta^{IJ}=\bar\theta^{IJ}-(x\cdot\gamma)\bar\epsilon^{IJ}.
\ee 
where $\bar\theta^{IJ}$ and $\bar\epsilon^{IJ}$ are constant spinors. The actual realization of the program relies of course on some educated guess on the structure of the couplings.
We discuss here the main ideas and show the explicit form of the relevant couplings. The structure of these loops and their quantum properties will be studied in greater detail in a future publication \cite{CGMS}.

 \section{Supersymmetry conditions for an arbitrary contour}

The  key  idea   exploited  in \cite{Drukker:2009hy}  to construct  1/2 BPS  lines and circles  is to embed the  natural $U(N)\times U(M)$  gauge  connection present in ABJ(M) theories   into  a super-connection\footnote{In Minkowski space-time, where $\psi$ and $\bar\psi$ are related by complex conjugation, $\mathcal{L}(\tau)$ belongs to $\mathfrak{u}(N|M)$ if $\bar\eta=i (\eta)^{\dagger}$. In Euclidean space, where the reality condition among spinors are lost,  we shall deal with the complexification  of this group $\mathfrak{sl}(N|M)$.}
\be
\label{superconnection}
 i\mathcal{L}(\tau) \equiv \begin{pmatrix}
i\mathcal{A}
&\sqrt{\frac{2\pi}{k}}  |\dot x | \eta_{I}\bar\psi^{I}\\
\sqrt{\frac{2\pi}{k}}   |\dot x | \psi_{I}\bar{\eta}^{I} &
i\hat{\mathcal{A}}
\end{pmatrix} \ \  \ \ \mathrm{with}\ \ \ \  \left\{\begin{matrix} \mathcal{A}\equiv A_{\mu} \dot x^{\mu}-\frac{2 \pi i}{k} |\dot x| M_{J}^{\ \ I} C_{I}\bar C^{J}\\
\\
\hat{\mathcal{A}}\equiv\hat  A_{\mu} \dot x^{\mu}-\frac{2 \pi i}{k} |\dot x| \hat M_{J}^{\ \ I} \bar C^{J} C_{I},
\end{matrix}\ \right.
\ee
belonging to the super-algebra of $U(N|M)$. In \eqref{superconnection} the coordinates $x^{\mu}(\tau)$  
describe the contour along which the loop operator is defined,  while  the quantities  $M_{J}^{\ \ I}(\tau)$, $\hat M_{J}^{\ \ I}(\tau)$, $\eta_{I}^{\alpha}(\tau)$ and $\bar{\eta}^{I}_{\alpha}(\tau)$  parameterize the possible
local couplings. The latter two, in particular, are taken to be Grassmann even quantities even though they transform in the spinor representation of the Lorentz group. We shall focus on operators  that possess
 a local $U(1)\times SU(3)$  $R-$symmetry invariance, since they are those described by semiclassical string surfaces in the dual picture. The $R-$symmetry structure of the couplings in \eqref{superconnection} is therefore described by a vector $n_{I}(\tau)$ (and its complex conjugate $\bar n_{I}$), that specifies the local  embedding of 
the unbroken $SU(3)$ subgroup into $SU(4)$\footnote{In the internal $R-$symmetry space $n_{I}$ identifies the direction preserved by the action of the  $SU(3)$ subgroup.}:
\be
\begin{split}
\label{cc}
&\eta_{I}^{\alpha} (\tau)=n_{I} (\tau)\eta^{\alpha} (\tau),\ \ \ \   \bar\eta^{I}_{\alpha} (\tau)=\bar n^{I} (\tau) \bar\eta_{\alpha} (\tau),\\     
M_{J}^{\ \ I} (\tau)=p_{1} (\tau)\delta^{I}_{ J}-&2 p_{2} (\tau) n_{J}  (\tau) \bar n^{I }(\tau),\ \ \ \ 
\widehat M_{J}^{\ \ I} (\tau)=q_{1}  (\tau)\delta^{I}_{J}-2 q_{2} (\tau) n_{J} (\tau) \bar n^{I} (\tau).
\end{split}
\ee
By rescaling the Grassmann even spinors $\eta^{\alpha}$ and $\bar\eta_{\alpha}$, we  can always choose $n_{I}\bar n^{I}=1$. The functions $p_{i}(\tau)$ and $q_{i}(\tau)$ appearing  in the definition of $M$ and $\hat M$ instead  control the eigenvalues of the two matrices.
The next step is to  constrain the form of the free  functions  present  in \eqref{cc} by requiring
 that the  Wilson loop defined by \eqref{superconnection} is globally supersymmetric. This  part of the construction 
 is quite different from its four-dimensional analog.  
The usual condition $\delta_{\rm susy}\mathcal{L}(\tau)=0$  is here too strong and it does not yield any solution for the  couplings \eqref{cc}.   To obtain non trivial  results, we must replace  $\delta_{\rm susy}\mathcal{L}(\tau)=0$  with  the weaker requirement \cite{Drukker:2009hy,Lee:2010hk}
\be
\label{var1}
\delta_{\rm susy}\mathcal{L}(\tau)=\mathfrak{D}_{\tau} G\equiv\partial_{\tau} G+ i\{ \mathcal{L},G],
\ee
where  the r.h.s. is the super-covariant derivative  constructed out of the  connection  
$\mathcal{L}(\tau)$  acting on a super-matrix $G$ in $\mathfrak{u}(N|M)$. The condition \eqref{var1} 
guarantees that the  action of the  supersymmetry charge translates into an infinitesimal $U(N|M)$
 super-gauge transformation for $\mathcal{L}(\tau)$ and thus   the {\it traced} loop operator is invariant\footnote{The stronger condition $\delta_{\rm susy}\mathcal{L}(\tau)=0$ would imply that both the super-holonomy and its {\it trace} are invariant. Here and in the text we are using the term {\it trace} loosely: 
 it might mean the actual trace, or the super-trace or something more exotic. The exact meaning will depend 
 on the specific family of Wilson loops we are considering.}.
 
 Since the supersymmetry  transformations of  the bosonic fields do not contain  derivatives,  the super-matrix $G$ in \eqref{var1} cannot have an arbitrary structure but it has to  be anti-diagonal, {\it i.e.} 
\begin{equation}
\label{susy2}
G=\begin{pmatrix}
0 & g_{1}\\
\bar g_{2}  & 0
\end{pmatrix}\ \ \Rightarrow \  \  \mathfrak{D}_{\tau} G=\begin{pmatrix}
\sqrt{\frac{2\pi}{k}}  |\dot x | (\eta_{I}\bar\psi^{I} \bar g_{2}-g_{1}\psi_{I}\bar \eta^{I}) &\mathcal {D}_{\tau} g_{1}\\
\mathcal{D}_{\tau }\bar g_{2}  & \sqrt{\frac{2\pi}{k}}  |\dot x | (-\bar g_{2}\eta_{I}\bar\psi^{I} +\psi_{I}\bar \eta^{I} g_{1}) 
\end{pmatrix} .
\end{equation}
Here the covariant derivative  $\mathcal{D}_{\tau}$ in \eqref{susy2} is constructed out of  the {\it dressed} bosonic connections $\mathcal{A}$ and $\hat{\mathcal{A}}$ and given by
\be
\begin{aligned}
\mathcal{D}_\tau g_{1}&= \partial_\tau  g_{1} + i (\mathcal{A}\,  g_{1} - g_{1}\, \hat {\mathcal{A}})\,, \ \ \ \ 
\mathcal{D}_\tau \bar g_{2} &= \partial_\tau \bar g_{2} 
- i (\bar g_{2}\,\mathcal{A}-   \hat{\mathcal{A}}\, {\bar g}_{2}) .\,
\end{aligned}
\ee
Supersymmetry is preserved if there exist two functions $g_{1}$ and $\bar g_{2}$ such that
%
\begin{subequations}
\label{cond1}
\begin{align}
\label{cond1a}
&\mathrm{(I)}:\ \ \  -i\sqrt{\frac{2\pi}{k}}|\dot{x}|\eta_I\delta\bar{\Psi}^I=\mathcal{D}_\tau g_1
&\mathrm{(II)}:\ \ \  -i\sqrt{\frac{2\pi}{k}}|\dot{x}|\delta\Psi_I\bar{\eta}^I=\mathcal{D}_\tau \bar{g}_2,\\
\label{cond1c}
&\mathrm{(III)}: \sqrt{\frac{2\pi}{k}}  |\dot x | (\eta_{I}\bar\psi^{I} \bar g_{2}-g_{1}\psi_{I}\bar \eta^{I})= \delta{\cal A},
&\mathrm{(IV)}:  \sqrt{\frac{2\pi}{k}}  |\dot x | (-\bar g_{2}\eta_{I}\bar\psi^{I}+\psi_{I}\bar \eta^{I}g_1)= \delta\bar{\cal A},
\end{align}
\end{subequations}
for a suitable form  of the couplings, taking into account the superconformal transformation of the ABJ(M) fields (see 
 \ref{susytransf}). The analysis can be performed in full generality and it will be presented in details in a forthcoming paper \cite{CGMS}: here we just state the main results, keeping track of their origin. 

First of all, the reduced spinor couplings $\eta_\alpha$ and $\bar\eta^\beta$  introduced in \eqref{cc} are determined by the contour $x^{\mu}$  through the relations
 \begin{align}
 \label{Bas/Eigen}
\mathrm{(A):}\ \ \ \  \delta^{\beta}_{\alpha}=  \frac{1}{2 i}(\eta^{\beta} \bar\eta_{\alpha}-\eta_{\alpha} \bar\eta^{\beta})\ \ \  \mathrm{and}\ \ \  
 \mathrm{(B):}\ \  \ \ \
{(\dot{x}^{\mu}\gamma_{\mu})_{\alpha}^{\ \ \beta}}=\frac{\ell}{2i} |\dot x|(\eta^{\beta} \bar\eta_{\alpha}+
\eta_{\alpha} \bar\eta^{\beta}).
 \end{align}
These conditions originate from (I) and (II) in (\ref{cond1a}), basically representing the request that derivative terms are taken along the contour. The matrices $M$ and $\widehat{M}$ have the form

\begin{equation}
M_{J}^{\ \ I} (\tau)=\widehat{M}_{J}^{\ \ I} (\tau)=\ell(\delta^{J}_{K}-2 n_{K}\bar n^{J}).
\end{equation}
The constant parameter $\ell$ can only take two values, $\pm 1$, and the choice specifies  the eigenvalues of the matrices $M_{J}^{\ \ I} (\tau)$ and $\widehat M_{J}^{\ \ I} (\tau)$: $(-1,1,1,1)$ [$\ell=1$] and $(1,-1,-1,-1)$ [$\ell=-1$]. The invariance of  \eqref{Bas/Eigen} under the replacement $(\eta,\bar\eta)\mapsto ( u\eta,{u}^{-1} \bar\eta)$  is instead related to (III) and (IV) in  (\ref{cond1c}), simply determining the relative scale of the reduced spinor couplings.

The $SU(4)$ tensor structure of the preserved supercharge $\bar\Theta^{IJ}$ is controlled by a couple of constraints, consisting of the following  algebraic relations
\begin{alignat}{4}
\label{2.30}
&\mathrm{(A):}\ \ \ \ \epsilon_{IJKL} (\eta\bar{\Theta}^{IJ})\bar n^{K}=0\ \ \ \ \ \ &\mathrm{and}\ \ \  \ \ \ \ \  &\mathrm{(B):}\ \  \ \ \
n_{I}(\bar\eta\bar\Theta^{IJ})=0,
\\
\intertext{where the vectors $n_{K}$ and $\bar n^{K}$ are defined in \eqref{cc}. 
Finally  there are   two sets of differential conditions  }
\label{ar}
&\mathrm{(A):}\ \ \ \ \bar{\Theta}^{IJ}\partial_{\tau} {\bar{\eta}}^K\epsilon_{IJKL}=0
\ \ \ \ \ \ &\mathrm{and}\ \ \  \ \ \ \ \  
&\mathrm{(B):}\ \  \ \ \
\bar\Theta^{IJ}\partial_{\tau}{\eta}_{I}=0.
\end{alignat}
They  ensure that the derivative term in the supersymmetry variation takes the correct form without leaving any unwanted remnant.  

We remark that all the above conditions are strictly local. To construct an actual supersymmetric Wilson loop, we must provide a family of couplings $(\eta, \bar\eta, n_{I}, \bar n^{I})$ so that the solution of the eqs. \eqref{2.30} and \eqref{ar} takes the form of a conformal Killing spinor,
{\it i.e.} the form \eqref{KillSpin}.
The  relations \eqref{Bas/Eigen}, \eqref{2.30} and \eqref{ar} provide a  complete set of supersymmetry conditions, but their form is not unique. For instance the requirement \eqref{2.30} is equivalent to the following expansion for the preserved supercharge in terms of  the couplings $\eta_{K}$ and $\bar\eta^{K}$:
  \be
 \label{susychar}
 \bar\Theta^{IJ}_{\alpha}=
 \frac{\ell}{2 i}\left[\bar \eta_{\alpha}^{I}\bar{h}^{J}-\bar{\eta}_{\alpha}^{J}\bar h^{I} -\mbox{$\frac{1}{2}$}\epsilon^{IJKL}
  \eta_{K\alpha} {m}_{L}\right],
 \ee
 where  the Grassmann odd vectors $m_{I}$ and $\bar h^{I}$ are defined from
 \be
\label{cond1suba}
\eta\gamma^\mu\bar{\Theta}^{KL}\ n_K= \frac{\dot{x}^\mu}{|\dot{x}|}  \bar h^{L},  \ \ \ \ \ 
-\epsilon_{IJKL}\bar n^{K} \bar\Theta^{IJ}\gamma^{\mu}\bar\eta=\frac{\dot{x}^\mu}{|\dot{x}|}   m_{L}
 \ee
 and they obey the orthonormality relations: $n_{I}\bar n^{I}=1,  n_{I} \bar h^{I}=\bar n^{I} m_{I}=0$. The explicit form of the gauge functions $g_1$ and $\bar{g}_2$ can be nicely written using these vectors
 
 \be
g_{1}\equiv 2\sqrt{\frac{2\pi}{k}}    (\bar{h}^{L} C_{L}), \ \ \ \ \  \bar g_{2}\equiv \sqrt{\frac{2\pi}{k}}  (m_{L} \bar{C}^{L}).
\ee

\section{Supersymmetric Wilson loops on $\mathbb{R}^3$}
Our first explicit construction concerns a family of Wilson loops of arbitrary shape, which preserve at least  
a supercharge  of the Poincar\`e type, {\it i.e.} a supercharge with $\bar\epsilon^{IJ}=0$. In this sense these
operators can be viewed as the three dimensional companion of the loops discussed by Zarembo in \cite{Zarembo:2002an}.
They can be also  considered a generalization of  the BPS straight-line constructed  by Drukker and Trancanelli in  \cite{Drukker:2009hy}, which is the simplest example enjoying this property.  
 
We start by considering the differential constraints \eqref{ar}, written in a way which is easier to solve:
\be
\partial_{\tau} \bar{h}^{L}+|\dot{x}| \eta\bar{\epsilon}^{KL}\  n_K=0,\ \ \ \ 
\partial_{\tau} m_{L}+|\dot{x}|\bar{n}^K(\bar{\epsilon}^{IJ}\bar\eta)\epsilon_{IJKL}=0.
 \ee 
For $\bar\epsilon^{IJ}=0$ the vectors $m_{I}$ and $\bar h^{I}$ are seen independent of  the  contour parameter $\tau$. To further proceed we contract (\ref{susychar}) and its dual with $\eta_\alpha$
\be
\label{cioni}
\eta\bar{\Theta}^{IJ}=\ell (\bar n^{I}\bar{h}^{J}-\bar{n}^{J}\bar h^{I})\ \ \ \ \mathrm{and}\ \ \ \ \ 
\epsilon_{IJKL} (\bar\eta\bar\Theta^{IJ})=\ell( n_{K} {m}_{L}-n_{L} {m}_{K})
\ee
and we observe that, for a generic contour,  these expansions are compatible with a constant $\bar\Theta^{IJ}$ if we choose, for instance, the following ansatz for $n_{I}$ and $\bar n_{I}$:
\be
\label{bnIZ}
 \bar n^{I}=(\eta\bar s^{I})
\ \  \ \ \ \mathrm{and}\ \ \ \ \
 n_{I}=(s_{I}\bar\eta). 
\ee
Here  $\bar s^{I}_{\alpha}$ are  four $\tau-$independent spinors and $\eta$ and $\bar\eta$ are determined by 
\eqref{Bas/Eigen}.  The normalisation condition $\bar n_{I} n^{I}=1$ is  equivalent to the following completeness relation on the spinors $ s^{\alpha}_{I}$ and $\bar s^{I}_{\alpha}$
 \be
 \label{dfg}
 \bar s^{I}_{\beta} s_{I}^{\alpha}=\frac{1}{2 i} \delta^{\alpha}_{\beta}.
 \ee
 We plug our ansatz into the algebraic conditions \eqref{2.30} and, after some work,  we can show that for a generic contour they are equivalent to the linear system of equations
\be
\label{zar1}
\epsilon_{IJKL}(\bar{\theta}^{IJ}\gamma_{\mu}\bar s^{K})=0\ \ \ \ \mathrm{and} \ \ \ \ \  (s_{I}\gamma_{\mu}\bar{\theta}^{IJ})=0.
\ee
The relations \eqref{zar1} also ensure that the remaining differential constraints \eqref{ar} are identically satisfied in the case of Poincar\`e charges. The general solution of the supersymmetry conditions \eqref{zar1} can be  written as follows 
\be
\label{piccolino}
\bar\Theta^{IJ\rho}=
\bar\theta^{IJ\rho}=\bar  v^{J}
\bar s^{I\rho}-\bar v^{I}
\bar s^{J\rho},\ \ \ \ \ \ \ \  {\rm with} \ \ \ \ \ \ \ \
 \bar v^{I} s_{I\beta}=0.
\ee
It is straightforward to check that the above ansatz solve the conditions \eqref{zar1}, in fact
\be
 (s_{I}\gamma_{\mu}\bar{\theta}^{IJ})= (s_{I}\gamma_{\mu}\bar s^{I}) \bar v^{J}\!-\!\bar v^{I}(s_{I}\gamma_{\mu}\bar s^{J}) =\frac{1}{2i} {\rm Tr}(\gamma_{\mu})\bar v^{J}\!=0\ \ \ \ \ \ 
\epsilon_{IJKL}(\bar{\theta}^{IJ}\gamma_{\mu}\bar s^{K})=2\epsilon_{IJKL}~\bar v^{J}(\bar s^{I}\gamma_{\mu}\bar s^{K})=0.
\ee
The first result follows from the completeness relation \eqref{dfg}, while the property $(\bar s^{I}\gamma_{\mu}\bar s^{K})=(\bar s^{K}\gamma_{\mu}\bar s^{I})$, which holds for bosonic spinors, is responsible for the second one. To show that any solution of \eqref{zar1} can be cast into the form \eqref{piccolino} requires  some more work and the detail of the proof will be given in \cite{CGMS}.
From  \eqref{piccolo} it also follows that these loop are generically $1/12$-BPS.

Summarizing we have constructed  a family of supersymmetric Wilson  loops of arbitrary shape, whose coupling are 
\begin{subequations}
\label{Zarcoup}
\begin{align}
\eta^{\alpha}_{I}=n_{I}\eta^{\alpha}=& s_{I}^{\beta}\bar\eta_{\beta}\eta^{\alpha}=is^{\beta}_{I}\left(\mathds{1}+\ell\frac{\dot{x}\cdot \gamma}{|\dot x|}\right)_{\beta}^{\ \ \alpha},
\ \ \ \ \  
\bar\eta_{\alpha}^{I}=\bar n_{I}\bar\eta_{\alpha}= \bar\eta_{\alpha}\eta^{\beta} \bar s^{I}_{\beta}=i\left(\mathds{1}+\ell\frac{\dot{x}\cdot \gamma}{|\dot x|}\right)_{\alpha}^{\ \ \beta}\bar s^{I}_{\beta},
\\
&M_{K}^{\ \ J} =\widehat M_{K}^{\ \ J} =\ell\left(\delta^{J}_{K}-2 n_{K}\bar n^{J}\right)
=\ell\left (\delta^{J}_{K}-2 i s_{K}\bar s^{J}-2 i\ell\frac{\dot x^{\mu}}{|\dot x|} s_{K}\gamma_{\mu}\bar s^{J} 
\right),
\end{align}
\end{subequations}
and which are invariant under the Poincar\`e  supercharges \eqref{piccolino}.

\section{Supersymmetric Wilson loops on $S^2$}
\label{sec:pert}

We propose a second family of Wilson loops, that is defined for an arbitrary curve on the unit sphere $S^{2}$: $x^{\mu} x_{\mu}=1$.   The central  idea in our construction is again a judicious guess for the reduced vector couplings $n_{I}$ and $\bar n_{I}$, which were introduced in \eqref{cc}. Specifically we shall consider a deformation of the ansatz \eqref{bnIZ} 
\be
\label{bnIS2}
 \bar n^{I}=r (\eta U\bar s^{I})
\ \  \ \ \ \mathrm{and}\ \ \ \ \
 n_{I}=\frac{1}{r} (s_{I}U^{-1}\bar\eta), 
\ee
where $ s^{I}_{\alpha}$  and $\bar  s^{I}_{\alpha}$ are   again  four  $\tau-$independent spinors obeying the completeness relation \eqref{dfg}. The parameter  $r$ is a  function of $\tau$ and it will become useful when we have to solve the differential constraints.  The matrix $U$ is the characterising ingredient of our ansatz:   it is an element of $SU(2)$ constructed with the coordinates $x^{\mu}(\tau)$ of the circuit, namely
\be
\label{Umatrix}
U=\cos\alpha~\mathds{1}+i \sin\alpha~ (x^{\mu}\gamma_{\mu}),
\ee   
with $\alpha$ free constant parameter.
There is a natural connection among $U$ in \eqref{Umatrix}, the tangent vector to the circuit and the invariant 
one-forms on $S^{2}$. In fact if we evaluate the Lie-algebra element $\partial_{\tau} U U^{\dagger}$, we obtain
\be
\begin{split}
\partial_{\tau} U U^{-1}
=&i \sin\alpha~(\cos\alpha~\dot x_{\lambda}-\sin\alpha~\epsilon_{\lambda\mu\nu} x^{\mu}\dot x^{\nu}) \gamma^{\lambda},
\end{split}
\ee
where the r.h.s. is a linear combination of the tangent vector and of the $SU(2)$ invariant forms.  

Let us first focus our attention on the algebraic conditions  (A) in \eqref{2.30}.  Using the Fierz identity and  the explicit form  \eqref{KillSpin} for the Killing spinor, we can rewrite it  as follows
 \begin{equation}
 \label{4.4}
\frac{r}{2}\epsilon_{IJMN}(\zeta \gamma^{\mu} \zeta) (\bar{\Delta}^{IJ} \gamma_{\mu}\bar s^{M})=0,
 \end{equation}
where we have defined an auxiliary  reduced coupling $\zeta= U^{-1}\eta$  and  an auxiliary super-conformal 
charge $\bar\Delta^{IJ}=\bar\Theta^{IJ}U$. We can also rearrange the  condition (B) in \eqref{2.30}  following the same idea and we find
\be
\label{4.5}
 \frac{1}{2r}(\bar{\zeta}\gamma^{\mu}\bar{\zeta})(s_{I}\gamma_{\mu}\bar{\Delta}^{IJ})=0,
\ee
where $\bar{\zeta}= U^{-1}\bar\eta$. Now we notice that eqs. \eqref{4.4} and  \eqref{4.5}, for a generic contour, leads to the same conditions (\ref{zar1}) discussed in the previous section\footnote{In the language of the previous section, these equation would arise for a circuit whose tangent vector is
\[
\dot{y}^{\mu}=\frac{|\dot x|}{2i}  \zeta\gamma^{\mu}\bar{\zeta}=
\frac{|\dot x|}{2i}  \eta U\gamma^{\mu} U^{-1}\bar{\eta}=\cos2\alpha \dot x^{\mu}+ \sin 2\alpha
\epsilon^{\mu\nu\lambda} x_{\nu}\dot{x}_{\lambda}.
\]}
\be
\label{cicap}
\epsilon_{IJKL}(\bar{\Delta}^{IJ}\gamma_{\mu}\bar s^{K})=0\ \ \ \ \mathrm{and} \ \ \ \ \  (\bar{\Delta}^{IJ}\gamma_{\mu} s_{I})=0.
\ee
Conversely they possess the same kind of solutions\footnote{Since $x^{2}=1$ it is straightforward to show that $\bar\Delta^{IJ}$ on $S^{2}$ has still the structure of a  conformal Killing spinor.}: $\bar\Delta^{IJ}$  is the  constant spinor  $\bar\theta^{IJ}$
defined in \eqref{piccolino}. In other words the preserved supercharges can be parametrized as follows
 \be
 \label{carsimp}
 \bar\Theta^{IJ}=[ \cos\alpha\, \mathds{1}+i \sin\alpha~ (x^{\mu}\gamma_{\mu})]
\bar\theta^{IJ}=U\bar\theta^{IJ}.
 \ee
The above  representation is very useful when we examine the derivative constraints \eqref{ar}: in fact it allows us to easily recognise  all the terms which  automatically vanish  since they are proportional  to the two SUSY conditions \eqref{zar1}.  Using this fact, the first of the two constraints \eqref{ar} can be easily translated into an ordinary differential equation for the unknown function $r$
\be
\dot{r}+i\ell r \sin\alpha\cos\alpha=0,
\ee
 which determines 
   the arbitrary function $r$:
   \be
   r= r_{0}\exp\left(-\frac{i}{2}\ell(\sin2\alpha) s\right).
   \ee
 Here $s$ is the affine parameter of the curve and $r_{0}$ an arbitrary constant. It is a simple exercise to show that the second differential constraint \eqref{ar} is identically satisfied. 
 
It is straightforward to compute the couplings  for this  family of supersymmetric Wilson  loops on $S^{2}$:
\begin{subequations}
\begin{align}
\eta^{\beta}_{I}=&
\frac{i}{r_{0}}e^{\frac{i}{2}\ell(\sin2\alpha) s}\left[s_{I}(\cos\alpha~\mathds{1}-i \sin\alpha~ (x^{\mu}\gamma_{\mu}))\left(\mathds{1}+\ell\frac{\dot{x}\cdot \gamma}{|\dot x|}\right)\right]^{\beta},
\\
\bar\eta_{\beta}^{I}=&
i r_{0}e^{-\frac{i}{2}\ell(\sin2\alpha) s}\left[\left(\mathds{1}+\ell\frac{\dot{x}\cdot \gamma}{|\dot x|}\right)\left(\cos\alpha~\mathds{1}+i \sin\alpha~ (x^{\mu}\gamma_{\mu})\right)\bar s^{I}\right]_{\beta},
\\
M_{K}^{\ \ J} =&\widehat M_{K}^{\ \ J} 
=\ell\left [\delta^{J}_{K}-2i s_{K}\bar s^{J}-2 i\ell\cos 2\alpha  \left(s_{K}\frac{\dot{x}\cdot \gamma}{|\dot x|} \bar s^{J}\right)-2 i \ell \sin2\alpha \left(s_{K}\gamma^{\lambda}\bar s^{J}\right)\epsilon_{\lambda\mu\nu} x^{\mu}\dot x^{\nu}
\right].
\end{align}
\end{subequations}
Some general remarks are now in order.  First of all we notice that for $\alpha=0$ we recover  the  couplings \eqref{Zarcoup}. In this sense we can consider  this class of loops as a deformation of those considered in the previous section. For generic $\alpha$, the situation is more intricate. Consider, for instance,  the structure of the  scalar couplings: there is a universal constant sector which  is not controlled by $\alpha$. Then we find a term of the form $R_{\mu K}^{\ \ \ J} \dot{x}^{\mu}$, which is the analog of  {\it Zarembo}  coupling in four dimensions. Finally  we have a  contribution of the type $T_{\ \ K}^{\lambda \ \ J} \epsilon_{\lambda\mu\nu} x^{\mu}\dot x^{\nu}$  describing  the  coupling of the scalars to the invariant forms on $S^{2}$. This is reminiscent of the  Wilson loops on $S^{3}$ in $D=4$ discussed  in \cite{Drukker:2007qr}.

In this picture the value $\alpha=0$ corresponds to the decoupling of the forms on $S^{2}$. There is a second interesting value of $\alpha$, {\it i.e.} $\alpha=\frac{\pi}{4}$, for which the {\it Zarembo-}like term vanishes and the scalars couple only to the invariant forms. For this value of $\alpha$ we also recover the $1/2$ BPS circle  discussed in \cite{Drukker:2009hy}. One is then tempted to identify these operators as the three dimensional companions of the so-called DGRT loops \cite{Drukker:2007qr}. 
\subsection{Gauge transformation and the construction of the invariant operator}
In order to construct a gauge-invariant operator we have to discuss the global effect of the super-gauge transformations related to supersymmetry: let us consider the infinitesimal super-gauge transformation that, in this case, affects the $S^2$ loops  
\be
G=\begin{pmatrix}
0 & g_{1}\\
\bar g_{2}  & 0
\end{pmatrix}
\ \ \mathrm{with}\ \   g_{1}\equiv 2\sqrt{\frac{2\pi}{k}}    (\eta_{I}\bar{\Theta}^{IL} C_{L})
\ \  \mathrm{and}\ \  \bar g_{2}\equiv \sqrt{\frac{2\pi}{k}}  ( \epsilon_{IJKL} (\bar\eta^{K}\bar\Theta^{IJ}) \bar{C}^{L}).
\ee
In general the functions $g_{1}$  and $\bar g_{2}$  for a closed loop are neither periodic nor anti-periodic. If  we  take the range of $\tau$  to be $[0,2\pi]$ and we denote with $L$ the perimeter of the curve, we find the following twisted boundary conditions
\be
g_{1}(2\pi)=g_{1}(0)e^{\frac{i}{2}(\sin2\alpha) L}\ \ \ \ \  \mathrm{and}\ \ \ \ \  \bar g_{2}(2\pi)=g_{2}(0)e^{-\frac{i}{2}(\sin2\alpha) L}.
\ee
Alternatively in matrix language we can write
\be
G(2\pi)=\begin{pmatrix} e^{\frac{i}{2}(\sin2\alpha) L}  & 0\\ 0 & e^{-\frac{i}{2}(\sin2\alpha) L} \end{pmatrix}G(0)=G(0)\begin{pmatrix} e^{-\frac{i}{2}(\sin2\alpha) L}  & 0\\ 0 & e^{\frac{i}{2}(\sin2\alpha) L} \end{pmatrix}.
\ee
If we introduce the auxiliary matrix 
\be
\mathcal{T}=\begin{pmatrix} e^{\frac{i}{4}(\sin2\alpha) L}  & 0\\ 0 & e^{-\frac{i}{4}(\sin2\alpha) L} \end{pmatrix},
\ee
we can easily show that the infinitesimal gauge transformation $G$  obey the following relation $G(2\pi)=\mathcal{T} G(0) \mathcal{T}^{-1}$, which in turn implies 
\be
U(2\pi)=\mathcal{T} U(0) \mathcal{T}^{-1},
\ee 
for the finite gauge transformation, $U=\exp(iG)$.
Then $\mathrm{STr}(\mathcal{W} \mathcal{T})$ defines a supersymmetric operator
\be
\mathrm{STr}(\mathcal{W} \mathcal{T})\mapsto \mathrm{STr}(U^{-1}(0)\mathcal{W} U(2\pi) \mathcal{T})= \mathrm{STr}(U^{-1}(0)\mathcal{W}\mathcal{T} \mathcal{T}^{-1} U(2\pi) \mathcal{T})=
\mathrm{STr}(\mathcal{W}\mathcal{T}).
\ee
In the case of the particular $\alpha=\frac{\pi}{4}$ and for the equatorial circle ($L=2\pi$), the twist matrix $\mathcal{T}$  is 
``$ i\sigma_{3}$'' which means that we have to  take the trace, as already shown in  \cite{Drukker:2009hy}.
The dependence of $\mathcal{T}$ on the perimeter of the curve is not  a complete surprise. In fact an 
hint of this result is implicitly  contained  in the original analysis of  \cite{Drukker:2009hy}  for the circle.
They suggest to use the trace since the gauge function are anti-periodic. However if we  cover the
circle twice (so doubling its length) the gauge functions are now periodic and thus we have to go back to 
the super-trace.
\subsection{An example: the great $\alpha$-circle}
As an example, we shall consider the great circle for generic $\alpha$
\be
\label{GreatCircle}
x^{1}=\cos\tau,\ \ \ \  x^{2}=\sin\tau,\ \ \ \  x^{3}=0.
\ee 
In this case the vector $\epsilon_{\lambda\mu\nu}x^{\mu}\dot{x}^{\nu}$ is $\tau-$independent and it is simply given by $(0,0,1)$.  In order to  write down explicitly the spinor and the scalar couplings for generic $\alpha$ it is convenient to introduce  the following parametrization  for the constant spinors $\bar s^{I}_{\alpha}$ and $s_{I\alpha}$
\be
\label{5.2}
\bar s^{I}_{\alpha}=\bar u^{I} \bar\lambda_{\alpha}+\bar v^{I} \lambda_{\alpha}\ \ \ \ \
\mathrm{and}\ \ \ \ \  s_{I\alpha}=  u_{I}\lambda_{\alpha}- v_{I}\bar\lambda_{\alpha},
\ee
where $\bar u^{I} u_{I}=\bar v^{I} v_{I}=1$  and $\bar u^{I} v_{I}=\bar v^{I} u_{I}=0$, while $\lambda$ and $\bar\lambda$ span a basis and they are normalised so that
\be
\lambda^{\alpha}\bar\lambda_{\beta}-\lambda_{\beta}\bar\lambda^{\alpha}=\frac{1}{2 i}\delta^{\alpha}_{\beta}.
\ee
For instance, we  can choose $\lambda$ and  $\bar\lambda$ to be  the eigenstate of $\gamma^{3}$ and in that case the couplings take the following form
\begin{subequations}
\begin{align}
\eta_{I}^{\alpha}=& \frac{i}{r_{0}}e^{\frac{i}{2}\ell(\sin2\alpha) \tau}\left[\cos\left(\alpha-\ell\frac{\pi}{4}\right)u_{I}- \sin\left(\alpha-\ell\frac{\pi}{4}\right)v_{I}
e^{i\tau}\right](1,-i\ell e^{-i\tau}),\\
\bar \eta^{I}_{\alpha}=&i r_{0}e^{-\frac{i}{2}\ell(\sin2\alpha) \tau}\left[\bar u^{I} \cos\left(\alpha-\ell\frac{\pi}{4}\right)-e^{-i\tau}\bar v^{I }\sin\left(\alpha-\ell\frac{\pi}{4}\right)
\right]\begin{pmatrix}-i \\ \ell e^{i\tau}
\end{pmatrix},
\end{align}
\begin{align}
&M_{K}^{\ \ J} (\tau)=\widehat M_{K}^{\ \ J} (\tau)
=\ell\left [\delta^{J}_{K}-2i s_{K}(1+\ell\sin 2\alpha\gamma^{3})\bar s^{J}-2 i\cos 2\alpha  \left(s_{K}\frac{\dot{x}\cdot \gamma}{|\dot x|} \bar s^{J}\right)
\right]=\\
&=\ell\left [\delta^{J}_{K}-\left((1+\ell \sin 2\alpha)
u_{K}\bar u^{J}+(1-\ell \sin 2\alpha)v_{K}\bar v^{J}\right)-\ell\cos 2\alpha  
(u_{K}\bar v^{J} e^{-i\tau}+v_{K} \bar u^{J}e^{i\tau} )
\right].\nonumber
\end{align}
\end{subequations}
A different choice for $\lambda$ and $\bar\lambda$  yields  equivalent coupling, which simply differs for a redefinition of $u_{I}$ and $v_{I}$.  For $\alpha=\pm\frac{\pi}{4}$, we obtain the well-known $1/2-$BPS circle of \cite{Drukker:2009hy}. 
In general we find 4 supercharges,
 {\it i.e.} the loops are  $1/6$-BPS.
 
 \noindent
 If we choose $x^{\mu}(\tau)$ to be two half-latitudes of the sphere differing of an angle $\delta$, we recover the supersymmetric wedge discussed in \cite{Griguolo:2012iq} for $\alpha= \frac{\pi}{4}$ and $\ell=1$. More examples and  other features  of these loops, such as their perturbative behaviour, will  be discussed in 
 \cite{CGMS}.
  
\section*{Acknowledgements}
This work was supported in part by the MIUR-PRIN contract 2009-KHZKRX. We warmly thank Flavio Porri for participating to the early stages of the these investigations.

\appendix
 \addcontentsline{toc}{section}{\large Appendices}
\renewcommand{\theequation}{\Alph{section}.\arabic{equation}}
\section{Spinor and supersymmetry transformations}
\label{susytransf}
In Euclidean space-time we choose the usual Pauli matrices as Dirac matrices: $\gamma^{\mu}\equiv\sigma^{\mu}$.  The spinor indices are raised and lowered as follows:  $\psi^{\alpha}=\epsilon^{\alpha\beta}
\psi_{\beta}$ and $\psi_{\alpha}=\epsilon_{\alpha\beta} \psi^{\beta}$ with $\epsilon^{01}=\epsilon_{10}=1$.

In ABJ(M) theories the gauge sector consists of  two  gauge fields $A_\mu$ and
$\hat{A}_\mu $ belonging respectively  to the adjoint of $U(N)$ and $U(M)$. The
matter sector  instead contains   the complex fields $C_I$ and ${\bar C}^{I}$ as well as the fermions $\psi_I$ and $\bar\psi^{I}$. The fields $(C,\bar \psi)$ transform in  the $({\bf N},{\bf \bar M})$  of the gauge group $U(N)\times U(M)$ while the couple $(\bar C, \psi)$ lives in the $({\bf \bar N},{\bf M})$. The additional  capitol index 
$I=1,2,3,4$  belongs to the $R-$symmetry group $SU(4)$.  Under a superconformal transformations defined by the  parameter $\bar{\Theta}^{IJ}_\alpha\equiv\bar{\theta}^{IJ}_\alpha+x_{\mu}\gamma^{\mu}_{\alpha\beta}\bar{\epsilon}^{IJ\beta}$ these fields transform as
\begin{align}
\delta A_\mu=&\frac{4\pi i}{k}\bar{\Theta}^{IJ\alpha}(\gamma_\mu)_\alpha^{\ \beta}\left(C_I\Psi_{J\beta}+\frac{1}{2}\epsilon_{IJKL}\bar{\Psi}_\beta^K\bar{C}^L\right)
\ \ \
\delta \hat A_\mu=\frac{4\pi i}{k}\bar{\Theta}^{IJ\alpha}(\gamma_\mu)_\alpha^{\ \beta}\left(\Psi_{J\beta}C_I+\frac{1}{2}\epsilon_{IJKL}\bar{C}^L\bar{\Psi}_\beta^K\right)
\nonumber\\
\delta C_K=&\bar{\Theta}^{IJ\alpha}\epsilon_{IJKL}\bar{\Psi}^L_\alpha
\ \ \ \ \ \ 
\delta \bar{C}^K=2\bar{\Theta}^{KL\alpha}\Psi_{L\alpha}
\nonumber \\
 \label{transf5}
\delta\Psi^\beta_K=&-i\bar{\epsilon}^{IL\beta}\epsilon_{ILKJ}\bar{C}^J
-i\bar{\Theta}^{IJ\alpha}\epsilon_{IJKL}(\gamma^\mu)_\alpha^{\ \beta}D_\mu \bar{C}^{L}\\
&+\frac{2\pi i}{k}\bar{\Theta}^{IJ\beta}\epsilon_{IJKL}(\bar{C}^LC_P\bar{C}^P-\bar{C}^PC_P\bar{C}^L)+\frac{4\pi i}{k}\bar{\Theta}^{IJ\beta}\epsilon_{IJML}\bar{C}^MC_K\bar{C}^L \nonumber\\
\delta\bar{\Psi}_\beta^K=&-2i\bar{\Theta}^{KL\alpha}(\gamma^\mu)_{\alpha\beta}D_\mu C_{L}-\frac{4\pi i}{k}\bar{\Theta}^{KL}_\beta(C_L\bar{C}^MC_M-C_M\bar{C}^MC_L)-\frac{8\pi i}{k}\bar{\Theta}^{IJ}_\beta C_I\bar{C}^KC_J-2i\bar{\epsilon}^{KL}_\beta C_L\nonumber
\end{align}


\end{document}